\documentclass[preprint,aps,floatfix,showpacs]{revtex4}

\usepackage{graphicx}
\usepackage{dcolumn}
\usepackage{bm}
\begin{document} 

\title{Measurements of nuclear $\gamma$-ray line emission 
in interactions of protons and $\alpha$ particles with N, O, Ne and Si.}

\author{H. Benhabiles-Mezhoud}
\affiliation{Centre de Spectrom\'etrie
Nucl\'eaire et de Spectrom\'etrie de Masse (CSNSM), CNRS-IN2P3 et Universit\'e
Paris-Sud, 91405 Orsay Campus, France}

\author{J. Kiener}   
\email{Jurgen.Kiener@csnsm.in2p3.fr} 
\affiliation{Centre de Spectrom\'etrie
Nucl\'eaire et de Spectrom\'etrie de Masse (CSNSM), CNRS-IN2P3 et Universit\'e
Paris-Sud, 91405 Orsay Campus, France}

\author{ J.-P. Thibaud}
\affiliation{Centre de Spectrom\'etrie Nucl\'eaire et de Spectrom\'etrie de
Masse (CSNSM),  CNRS-IN2P3 et Universit\'e Paris-Sud,
91405 Orsay Campus, France}

\author{ V. Tatischeff} 
\affiliation{Centre de Spectrom\'etrie Nucl\'eaire et de Spectrom\'etrie de
Masse (CSNSM),  CNRS-IN2P3 et Universit\'e Paris-Sud,
91405 Orsay Campus, France}

\author{I. Deloncle}   
\affiliation{Centre de Spectrom\'etrie Nucl\'eaire et de
Spectrom\'etrie de Masse (CSNSM), CNRS-IN2P3 et Universit\'e Paris-Sud, 91405 Orsay
Campus, France}

\author{A. Coc}  
\affiliation{Centre de Spectrom\'etrie Nucl\'eaire et de
Spectrom\'etrie de Masse (CSNSM), CNRS-IN2P3 et Universit\'e Paris-Sud, 91405 Orsay
Campus, France}

\author{ J.-C. Dalouzy}
\affiliation{GANIL, CEA/DSM-CNRS/IN2P3, Caen, France}

\author{ F. Dayras}
\affiliation{CEA, LIST, 91191 Gif-sur-Yvette CEDEX, France}

\author{ F. de Grancey}
\affiliation{GANIL, CEA/DSM-CNRS/IN2P3, Caen, France}

\author{ F. de Oliveira}
\affiliation{GANIL, CEA/DSM-CNRS/IN2P3, Caen, France}

\author{ N. de S\'er\'eville}
\affiliation{Institut de Physique Nucl\'eaire (IPN), CNRS-IN2P3 et Universit\'e
Paris-Sud, 91400 Orsay, France}

\author{J. Duprat}   
\affiliation{Centre de Spectrom\'etrie Nucl\'eaire et de
Spectrom\'etrie de Masse (CSNSM), CNRS-IN2P3 et Universit\'e Paris-Sud, 91405 Orsay
Campus, France}

\author{C. Hamadache} 
\affiliation{Centre de Spectrom\'etrie Nucl\'eaire et de
Spectrom\'etrie de Masse (CSNSM), CNRS-IN2P3 et Universit\'e Paris-Sud, 91405 Orsay
Campus, France}

\author{L. Lamia}
\affiliation{INFN - Laboratori Nazionali del Sud, via S. Sofia 62, 95125 Catania, Italy }

\author{A. Lefebvre-Schuhl} 
\affiliation{Centre de Spectrom\'etrie Nucl\'eaire
et de Spectrom\'etrie de Masse (CSNSM), CNRS-IN2P3 et Universit\'e Paris-Sud, 
91405 Orsay Campus, France}

\author{S. Ouichaoui} \affiliation{USTHB, Facult\'e de Physique, BP 32, El-Alia,
16111 Bab Ezzouar, Algiers, Algeria}

\author{M.-G. Pellegriti} 
\altaffiliation[present address:]{INFN - Laboratori Nazionali del Sud, via S. Sofia 62, 95125 Catania, Italy }
\affiliation{Institut de Physique Nucl\'eaire (IPN), CNRS-IN2P3 et Universit\'e
Paris-Sud, 91400 Orsay, France}
  
\date{\today}

\begin{abstract} $\gamma$-ray production cross sections have been measured in proton irradiations of N, Ne and Si and $\alpha$-particle irradiations of N and Ne. In the same experiment we extracted also line shapes for strong $\gamma$-ray lines of $^{16}$O produced in proton and $\alpha$-particle irradiations of O. For the measurements gas targets were used for N, O and Ne and a thick foil was used for Si. All targets were of natural isotopic composition. Beams in the energy range up to 26 MeV for protons and 39 MeV for $\alpha$-particles have been delivered by the IPN-Orsay tandem accelerator. The $\gamma$ rays have been detected with four HP-Ge detectors in the angular range 30$^{\circ}$ to 135$^{\circ}$. We extracted 36 cross section excitation functions for proton reactions and 14 for $\alpha$-particle reactions. For the majority of the excitation functions no other data exist to our knowledge. Where comparison with existing data was possible usually a very good agreement was found. It is shown that these data are very interesting for constraining nuclear reaction models. In particular the agreement of cross section calculations in the nuclear reaction code TALYS with the measured data could be improved  by adjusting the coupling schemes of collective levels in the target nuclei $^{14}$N, $^{20,22}$Ne and $^{28}$Si. The importance of these results for the modeling of nuclear $\gamma$-ray line emission in astrophysical sites is discussed.

\end{abstract}


\maketitle

\section{Introduction}

Nuclear $\gamma$-ray line emission induced by interactions of energetic particles with the solar atmosphere is commonly observed from eruptive solar events like solar flares. The bulk of this emission is due to reactions of accelerated protons, $^3$He and $\alpha$ particles with abundant nuclei of the solar atmosphere and reactions of energetic heavy ions with ambient hydrogen and helium. In both populations, ambient and energetic, the most important nuclei heavier than He are usually the major isotopes of elements C, N, O, Ne, Mg, Si and Fe. De-excitation lines of all those nuclei (except $^{14}$N) have already been observed from solar flares. The intensities of prominent narrow lines induced by energetic light ions allow determination of ambient abundances and of the composition and energy spectra of accelerated light particles \cite{MRK, RMKM, SM95,SM98}. A thorough line-shape analysis may furthermore reveal directional distributions of the accelerated particles \cite{Smith,MKR,SMKS,SM03,Ofl}.

The same prominent lines are also expected from energetic heavy ions interacting with ambient H and He, but are much broader because  of the relatively high velocities of the emitting nuclei. Their analysis is, however, much more difficult because the broad lines are merged into a quasi-continuum component of numerous unresolved weak lines. A successful extraction of these lines would inform about the composition, energy spectra and directional distribution of the accelerated heavy-ion component \cite{SMgms}. 

Similar $\gamma$-ray line emissions featuring the same prominent narrow and broad lines as well as the weak-line quasi-continuum are expected from the interaction of galactic cosmic rays with the interstellar gas and dust \cite{RKL}. Such an emission is probably dominated by the low-energy part of the cosmic-ray energy spectrum. Searches for nuclear $\gamma$-ray lines from nearby molecular clouds like the Orion Molecular Cloud or the galactic center and galactic thin disk have not been successful up to date. Their eventual detection would bring enormous progress for our knowledge of the low-energy cosmic-ray component. At the moment, practically nothing is known about the cosmic-ray composition and spectrum below a few hundred MeV per nucleon because these particles are effectively deflected by magnetic fields which are carried by  the outstreaming solar wind and do not reach the inner heliosphere.

Interpretation of observed $\gamma$-ray spectra requires a large amount of nuclear cross section data. For line intensities, $\gamma$-ray production cross sections are needed from reaction threshold to typically hundred MeV per nucleon for solar flares and to hundreds of MeV per nucleon for cosmic-ray interactions. To this purpose a database has been built, initiated in 1979 by Ramaty, Kozlovsky and Lingenfelter \cite{RKL} and successively updated \cite{KMR, MKKS}. The last version \cite{MKKS} contains cross section excitation functions for the production of 181 different $\gamma$-ray lines, essentially from proton and $\alpha$-particle reactions with nuclei from He to Fe and from some $^3$He-induced reactions. The excitation functions cover the range from reaction threshold up to typically several hundred MeV per nucleon. 

For most of these lines experimental cross sections exist at projectile energies  below 25 MeV for protons and 40 MeV for $\alpha$ particles. Many of them were measured at the tandem accelerators in Washington \cite{Dyerp, Dyera, Seam} and Orsay \cite{OT, tanhe3, tan2002}. For some lines cross sections are available at higher energies from cyclotron laboratories \cite{Zobel, Naraya, Lang, Lesko}. Cross section interpolations and extrapolations to higher energies are based on nuclear reaction systematics and calculations with nuclear reaction codes like TALYS \cite{talys}. All other, presumably, weaker lines not explicitely listed in \cite{MKKS} and probably to be counted in tens of thousands are supposed to form the above mentionend quasi-continuum. For this component cross sections in \cite{MKKS} are entirely from TALYS predictions. 

We decided to extend the database for proton and $\alpha$-particle reactions with two abundant elements of the solar atmosphere N and Ne, for which $\gamma$-ray production cross sections are only available for a few lines. Reactions with neon make probably an important contribution to the weak-line quasi-continuum in solar flares and in cosmic-ray interactions. Nitrogen is also very important for the understanding of the $\gamma$-ray emission of Earth's atmosphere induced by interactions of cosmic rays and solar energetic particles. Furthermore we measured the $\gamma$-ray production from proton reactions with Si at three energies in order to extend the cross section database and test the nuclear reaction calculations for this element which is very important for the weak-line quasi-continuum component.

Line shape modeling needs differential cross sections for the emitting nucleus, the $\gamma$-ray and  correlations between them. At relatively low projectile energies ($\leq$ $\sim$ 15 MeV), compound nucleus resonances and interferences with the direct reaction mechanism make nuclear reaction calculations very difficult. Simultaneous fits of measured line shapes at different laboratory angles to obtain a parameter set for line-shape calculations is in this situation probably the only alternative. Up to now this has been done in detail only for the 4.438-MeV line of $^{12}$C \cite{RKL, lshape} and has been subsequently used to analyse this line in the $\gamma$-ray spectra of solar flares \cite{SMKS, Ofl, Harris}.

Therefore we decided to measure the line shapes for another nuclear de-excitation line, the 6.129-MeV line of $^{16}$O. This line has been chosen for two reasons: (1) it is usually one of the strongest de-excitation lines in solar flares; (2) $^{16}$O has only few $\gamma$-ray emitting levels which reduces the importance of $\gamma$-ray cascades and facilitates the line-shape calculations.  

The experiment is described in section II; obtained results for $\gamma$-ray line shapes and production cross sections are reported in section III where they are compared with previously reported data. A comparison of experimental data to nuclear reaction code calculations is presented in section IV.

\section{Experiment}

The experiment has been done at the tandem Van-de-Graaff accelerator of the IPN Orsay. Proton beams in the energy range 6.55-26.2 MeV and $\alpha$-particle beams in the range 7.5-39.2 MeV passed through a gas cell containing the target elements Ne, N, and O, where all employed gases were of natural isotopic composition. Three different molecular gases were used for the irradiation of N and O. A N$_2$O gas target was used for studying  lines produced by reactions of low-energy proton and $\alpha$-particle beams with $^{14}$N and $^{16}$O. At energies where spallation reactions of $^{16}$O may produce $^{14}$N lines, N$_2$ and CO$_2$ were used separately for measurements on $^{14}$N and $^{16}$O, respectively. For the irradiations of Si with proton beams of 10, 15 and 20 MeV we used a 21.1 mg/cm$^2$ thick foil of Si of natural isotopic composition placed at the center of the empty gas cell. 

The gas cell has been built as a 13 cm long aluminium cylinder of 3.8 cm inner diameter. Two narrower sections were placed on both ends of the cylinder, each of them holding a 6 $\mu$m thick Mylar foil. These sections had 12 and 20 mm diameter at the beam entrance and exit, respectively. The total length of the cell between the two foils was 18.7 cm. Three gas bottles were connected to the cell by individual gas dosing valves and an oil-free vacuum pump was connected by a standard manual valve. For each irradiation run the gas cell was filled by manual operation with the required gas and pressure. The gas pressure was chosen following a compromise between reasonably high $\gamma$-ray yield and small energy loss and angular straggling. It was measured with a piezo-electric vacuum gauge with an acurracy of better than 0.25\% and cell temperature was determined to better than 0.5 $^{circ}$C. Both values were noted before and after each irradiation run. Typical gas column densities and energy losses of the projectile for the different irradiation runs are listed in Table \ref{table_runs}. 

A Faraday cup was placed approximately 1.5 m downstream of the gas cell for the beam current determination. It consisted of an electrically isolated stainless steel tube of 75 cm length with an end cap of Al. A thick Pb absorber was placed around the Faraday cup to reduce the $\gamma$-ray and neutron fluxes originating in the cup towards the detection setup. Typical beam currents were 1-4 nA and typical irradiation times 15-30 minutes. We estimate the uncertainty of the beam charge determination to less than 5\%. 

Beam position and size were checked for each new energy by means of beam-induced fluorescence of an alumina target situated approximately 1 m upstream of the gas target. The typical beam spot size was less than 5 mm diameter. A further fine tuning of the beam was achieved by minimizing the $\gamma$-ray lines of $^{27}$Al at 843, 1014 and 2211 keV which were probably due to interactions between the beam and the sections holding the Mylar foils. A further check of beam transmission through the gas cell was possible by comparison of the beam currents in a Faraday cup upstream and the Faraday cup downstream of the gas cell. Both currents agreed typically to better than 5\%. For each beam energy we made a run with the gas cell being empty (P $\leq$ 0.1 mbar) in order to measure the beam-induced $\gamma$-ray background component which is dominated by interactions in the Mylar foils and the Faraday end cap. We made also several runs without beam to determine the room background.  

$\gamma$ rays were detected with four Ge detectors equipped with BGO shields for Compton suppression. Two segmented clover detectors VEGA from the Euroball setup \cite{VEGA} were placed at angles of 120$^{\circ}$ and 135$^{\circ}$ with respect to the beam direction and at 40 cm from the centre of the gas cell. Another clover detector - a prototype of the Exogam setup \cite{Exogam} - and a single coaxial Ge detector from the Eurogam phase I setup \cite{Eurogam} were situated at 90$^{\circ}$ and 20 cm and at 30$^{\circ}$ and 17.5 cm from the gas target, respectively. We took data for the 13 individual crystals with standard NIM electronics and the data acquisition system NARVAL \cite{NARVAL}. The count rates for each crystal were below 15 kHz and the corresponding acquisition dead times smaller than 25\%. Note that the dead-time correction is done for each crystal separately, which is a specific feature of the used acquisition system NARVAL \cite{NARVAL}. Uncertainties on acquisition dead times were estimated in the analyses of isotropic $\gamma$-ray angular distributions to be less than 3\%. 

\begin{figure}
\includegraphics[height=.6\textheight]{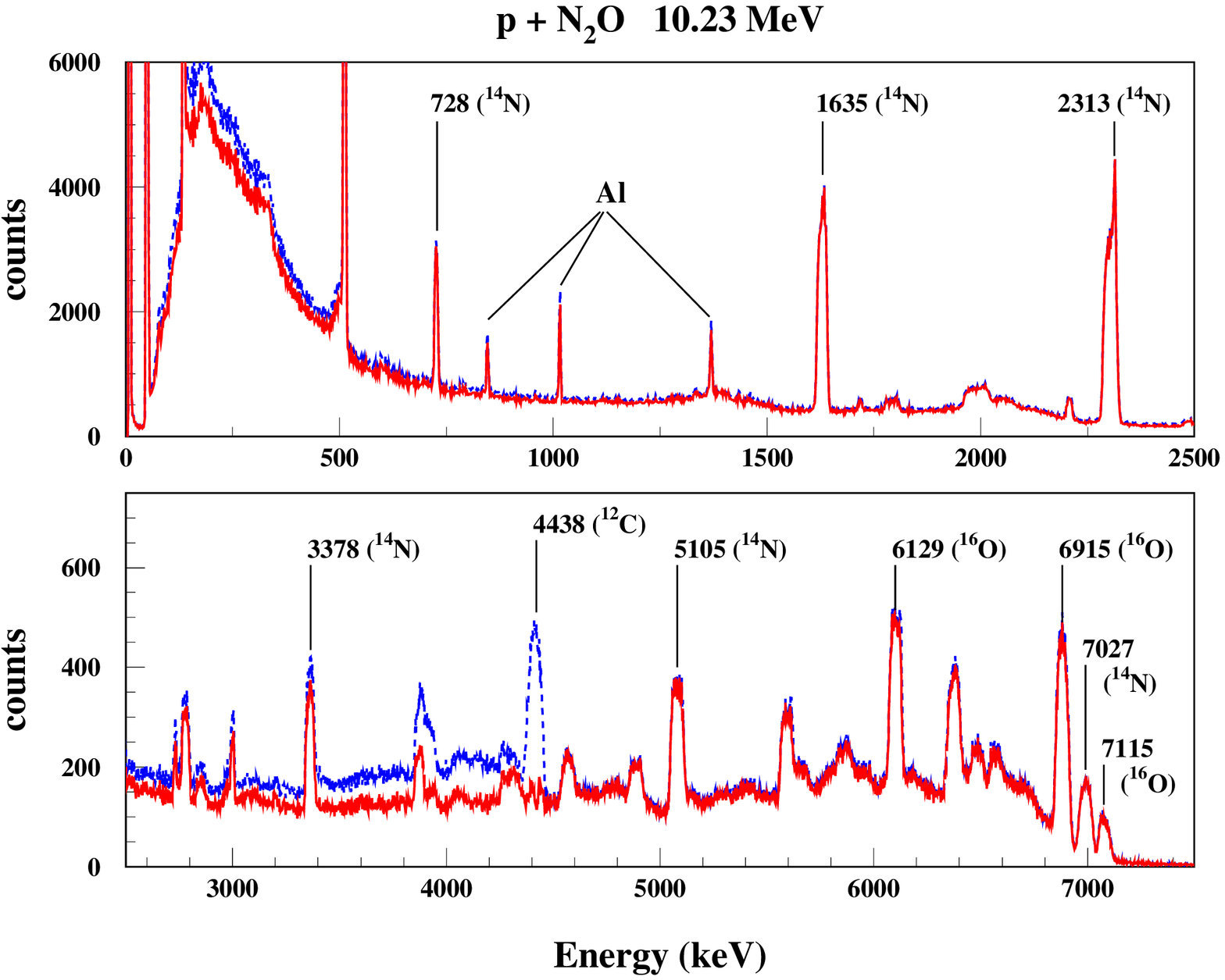}
\caption{(Color online) $\gamma$-ray spectra from 10.23 MeV proton irradiations obtained with the VEGA clover detector at 120$^{\circ}$. The dashed blue line shows the spectrum of the irradiation of the gas cell filled with N$_2$O, the full red line the spectrum obtained after subtraction of the empty-cell irradiation run. Both spectra are averages of spectra of the four individual crystals. The origin of the most prominent lines above the 511-keV line is indicated, ''Al'' standing for background lines produced by interactions with the surrounding material, mostly composed of Al.}
\label{figspec}
\end{figure}

Figure \ref{figspec} shows a typical spectrum for proton irradiation of N$_2$O at E$_p$ = 10.23 MeV. At this energy the 4.438-MeV line of $^{12}$C can essentially only be produced by proton inelastic scattering off $^{12}$C present in the Mylar foils, proton-induced reactions on $^{14}$N or $^{16}$O feeding the first excited state of $^{12}$C being ineffective. After subtraction of the empty-gas-cell run the spectrum is (beneath a strong 511-keV line) largely dominated by lines from $^{14}$N and $^{16}$O. Some relatively weak background lines are also observed, most probably from proton interactions with the Al tube of the narrow section at the gas cell exit. As these lines are much weaker in the empty gas cell runs and the beam optics remained unchanged this background is certainly due to angular straggling in the gas. Neutron interactions within material close to the detectors which is essentially made of Al may also contribute, where neutrons are essentially from the (p,n) reaction on $^{14}$N. 

The efficiency calibration of the detection setup was done with the help of Geant \cite{Geant} simulations and radioactive sources of $^{152}$Eu and $^{60}$Co. Both sources were successively placed at five different positions on the gas cell axis covering the beam path inside the gas. These measurements in the range E$_{\gamma}$ = 0.2-1.4 MeV were used to normalize the efficiency curves in the range 0.2-10 MeV obtained in detailed Geant simulations of the detection setup. The efficiency of the experimental setup was then taken as the average of the normalized efficiency curves for the five different positions on the gas cell axis. A more detailed description of the experiment and calibrations can be found in \cite{Benh}.

\section{Results: line shapes and cross sections}

All results have been obtained with the Compton-suppressed spectra of the individual crystals of the three clover and the single coaxial detector. We subtracted systematically the spectra resulting from the irradiation runs with empty gas cell with a  normalization following the dead-time corrected beam charges. An accurate subtraction was particularly important for irradiations where the same lines were produced in both the gas and the Mylar foils. Due to the C and O content of Mylar (chemical composition C$_{10}$H$_8$O$_4$), it concerned all measurements of the 6.129-MeV line shape and the measurements of $^{14}$N lines at projectile energies where fusion-evaporation reactions with $^{12}$C and $^{16}$O can populate excited states in $^{14}$N.

The total uncertainty in the subtraction was estimated to be less than 10\%, composed of a combined uncertainty of 7.5\% for the two charge measurements of the runs with and without gas and less than 3\% for the dead-time determination. An independent check of this uncertainty was possible by comparing the intensities of the 4.438-MeV line in the irradiation runs at low projectile energies. In fact this line was produced in the runs below E$_p$ $\approx$ 10 MeV and E$_{\alpha}$ $\approx$ 15 MeV exclusively by inelastic scattering off $^{12}$C present in the Mylar foils, because fusion-evaporation reactions with $^{14}$N and $^{16}$O of the N$_2$O gas populating the first excited state of $^{12}$C are energetically forbidden. The relative intensities of the 4.438-MeV line agreed typically to within 10\% with the relative beam charges of the runs with and without gas. 

At higher projectile energies the 4.438-MeV line is also produced in the N$_2$ and CO$_2$ gases by reactions with $^{14}$N and with $^{12}$C and $^{16}$O, respectively, but we extracted neither line shapes nor cross sections for it. For the latter two isotopes, a comprehensive data set of high-statistics line shapes and cross sections with probably better accuracy than could have been extracted from this experiment exists already. In reactions with $^{14}$N production cross sections for the 4.438-MeV line are much smaller than for inelastic scattering off $^{12}$C, such that the subtraction introduced an important uncertainty at many projectile energies, not allowing for a reasonable determination of the excitation function. 

The intensities of $\gamma$ rays produced by proton interactions with oxygen of the Mylar foils were generally much smaller than the ones for interactions within the gas, due to the small surface density of the Mylar foils with respect to the gas. A typical situation can be seen for the spectrum on fig. \ref{figspec} where the contribution of the Mylar foil is remarkable only for the 4.438-MeV line. Due to the lower gas pressures in the $\alpha$-particle runs, the contribution of the Mylar foils to the $^{16}$O lines was more important and reached up to 60\% for runs with N$_2$O gas at low $\alpha$-particle energies. This affected the 6.129-MeV line shape, but not the cross section determination for the $^{14}$N lines because neither the $^{12}$C($\alpha$,x)$^{14}$N nor the $^{16}$O($\alpha$,x)$^{14}$N reaction channels were open. 

\subsection{The 6.129-MeV line shapes}

Considering typical abundances and cross sections the 6.129-MeV line is in astrophysical sources like solar flares essentially produced by inelastic scattering off $^{16}$O. For harder energetic-particle spectra like galactic cosmic rays spallation of $^{20}$Ne may play a small role on the order of 10\%. Other reactions like $\alpha$-particle induced fusion-evaporation reactions with $^{14}$N are in most cases completely negligible. 

Two models for the calculation of the 6.129-MeV line profiles in astrophysical sources, in  particular solar flares, have been published: first, astrophysically motivated models applicable to general inelastic scattering reactions have been proposed in \cite{RaCr, RKL, MKR}. They employed energy-dependent few-parameter descriptions of the scattering and $\gamma$-ray emission process.  Later on \cite{lshape} and \cite{Werntz} developed models based on optical-model calculations for the 4.438 and 6.129-MeV line shapes. This was used for the analysis of the spectra of 19 solar flares observed by SMM \cite{SMKS} and of three solar flares observed by the high-resolution Ge spectrometer SPI onboard Integral \cite{Ofl, Harris}. In all models, however, the complete lack of measured line shapes for the 6.129-MeV line introduced a substantial uncertainty in the calculations. This should be especially important at low energies where individual compound-nucleus resonances dominate the inelastic scattering reactions which are certainly not well described by optical-model calculations or general parameterizations.

We extracted line shapes for the most important reaction p + $^{16}$O for 15 different proton energies in the range E$_p$ = 6.55 - 12.83 MeV with small steps of 0.3 - 0.5 MeV. Beam energies and gas densities were chosen to obtain a neat coverage of the complete energy range when considering the energy losses in the gas (see table \ref{table_runs}). In the range E$_p$ = 13.5 - 26.2 MeV, isolated compound nucleus resonances are less important and 6 different proton energies with larger energy steps of 1.5 - 3.7 MeV appeared sufficient to check the optical model calculations.  A similar energy coverage was used for $\alpha$-particle reactions with $^{16}$O: small energy steps of 0.5 MeV and neat coverage for E$_{\alpha}$ = 7.0 - 10 MeV and larger steps of 1 - 5 MeV up to 39.2 MeV.

In order to increase statistics, we created three new spectra by summing the individual background-subtracted spectra of the four crystals of each clover detector. This resulted in four spectra for each irradiation run, with at least two line profiles of good statistics (typically more than 10000 counts) from the VEGA clover detectors. Spectra of the single coaxial and the EXOGAM clover detector had lower statistics due to lower detection efficiency at 6 MeV. We then subtracted the background below the 6.129-MeV line which  was essentially due to remaining Compton scattering events of higher-energy $\gamma$-ray lines (6.915 and 7.115 MeV of $^{16}$O and 6.445 and 7.027 MeV of $^{14}$N) which were not suppressed by a veto signal of the BGO shields. For the subtraction we used the results of the Geant simulations of the detection setup.

\begin{figure}
\includegraphics[height=.6\textheight]{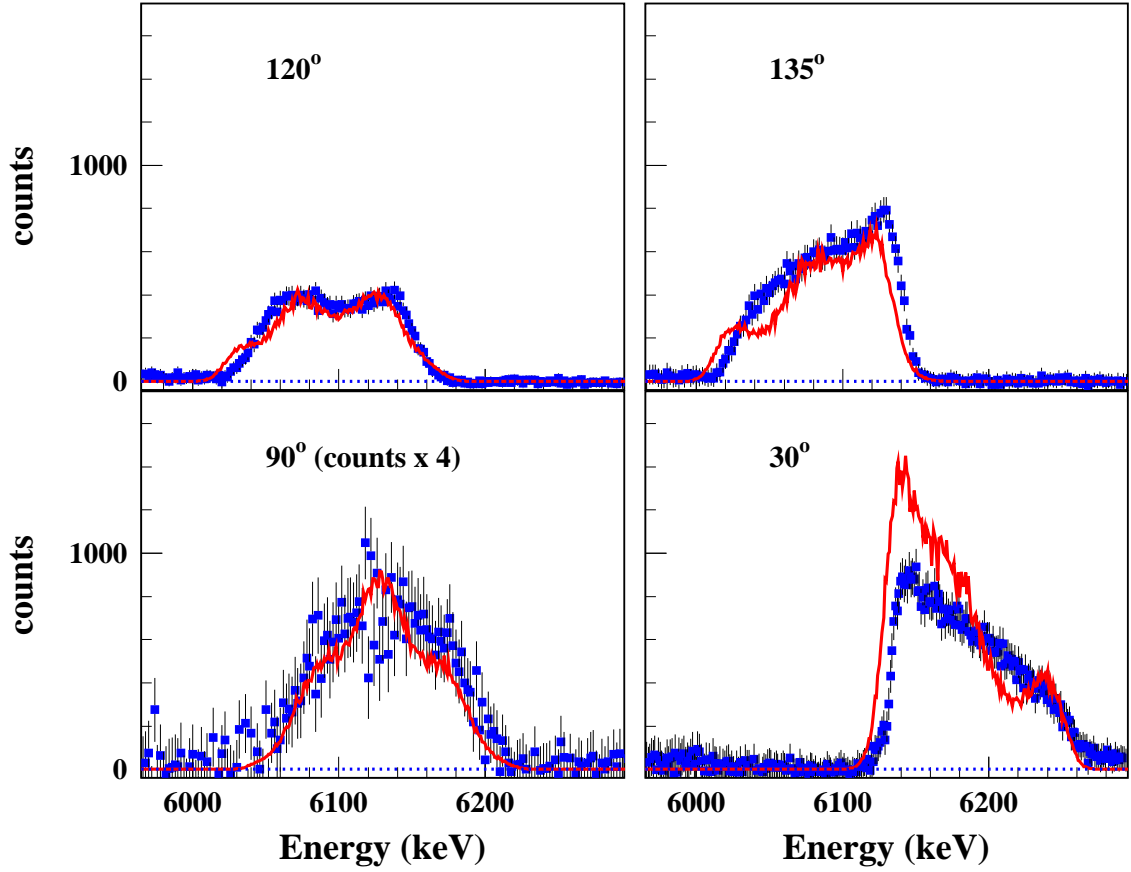}
\caption{(Color online) Profiles of the 6.129 $\gamma$-ray line of $^{16}$O obtained in the proton irradiation of N$_2$O gas at E$_p$ = 20 MeV at the indicated detection angles (blue squares). Data at 120$^{\circ}$, 135$^{\circ}$ and 90$^{\circ}$ are beam-induced background- and Compton-subtracted averages of the individual crystals of the clover detectors and at 30$^{\circ}$ of the single coaxial detector. The full red line is the result based on optical-model calculations in the coupled-channels formalism (see \cite{Ofl}), the dotted blue one represents the zero-baseline. The data at 90$^{\circ}$ have been multiplied by 4 for better visibility.}
\label{prof_20MeV} 
\end{figure}

\begin{figure}
\includegraphics[height=.6\textheight]{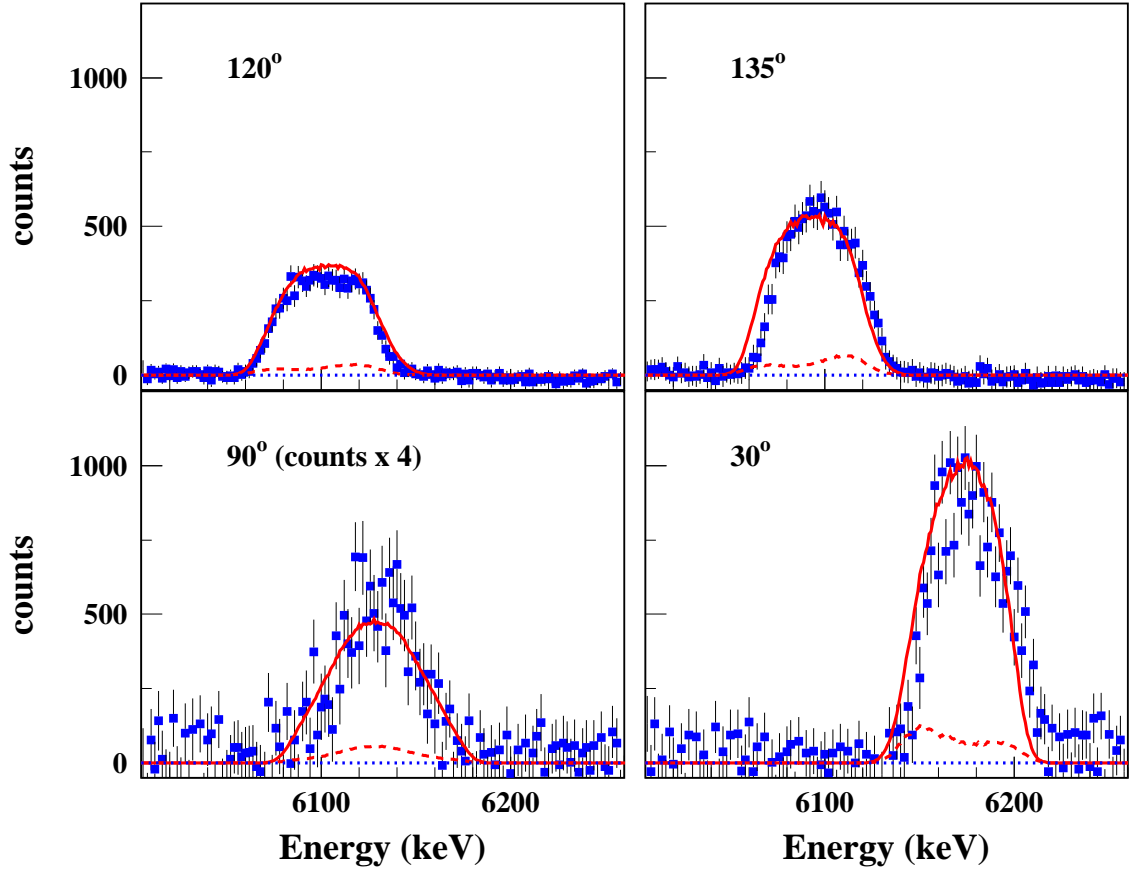}
\caption{(Color online) Data as in fig. \ref{prof_20MeV} except for E$_p$ = 10.23 MeV. The full red line is the sum of profiles from the direct reaction mechanism based on optical-model calculations (10\% of the integrated cross section) and a compound nucleus resonance with J$^{\pi}$ = 5/2$^+$. The dashed red line shows the contribution of the direct reaction mechanism alone.}
\label{prof_10MeV} 
\end{figure}

Examples of 6.129-MeV line shapes in proton reactions and results of calculations are shown in figs.  \ref{prof_20MeV} and \ref{prof_10MeV}. At E$_p$ = 20 MeV one expects the direct reaction mechanism to dominate inelastic scattering to the J$^{\pi}$=3$^-$, 6.130-MeV state of $^{16}$O. Optical model calculations should then reasonably well reproduce the $\gamma$-ray angular distributions and line shapes. An example is shown on fig. \ref{prof_20MeV}, where the line shape calculation has been done with the same coupled-channels calculations which were used for the analysis of the solar flares observed by SPI/INTEGRAL \cite{Ofl, Harris}. A good reproduction of the relative intensities and line shapes at the four detection angles is observed, except maybe at 30$^{\circ}$. 

Again, as expected, the optical model calculations from \cite{Ofl} fail to reproduce the line shapes at low energies. This is illustrated on fig. \ref{prof_10MeV} for E$_p$ = 10.23 MeV where the predictions of the direct reaction mechanism alone are clearly incompatible with the data. At this projectile energy several known resonances in $^{17}$F can be populated which may contribute to the 6.129-MeV line production. We finally found a good reproduction of the relative line intensities and the shapes assuming a pure 5/2$^+$ compound-nucleus resonance in $^{17}$F decaying to $^{16}$O$^{\star}$(J$^{\pi}$=3$^-$, 6.130 MeV) by l=1 proton emission with a possible small contribution of the direct reaction mechanism of the order of 10\%.

The explicit inclusion of compound-nucleus resonances at low energies improves here considerably the quality of line-shape calculations which has also been observed in the case of the 4.438 MeV $^{12}$C line \cite{Roorkee}. Similar findings hold also for the 6.129-MeV line in $\alpha$-particle inelastic scattering. A complete description of line shapes in inelastic scattering, however, including resonances and direct reactions plus eventually the contribution of spallation reactions is out of the scope of this paper. It will be presented elsewhere \cite{nlshape} including also a reevaluation of the 4.438-MeV line for which we have new data from threshold to 25 MeV for proton reactions and to 37.5 MeV for $\alpha$-particle reactions \cite{OT,tan2002}.

\subsection{Cross sections}

Differential cross sections for $\gamma$-ray lines emitted in the proton and $\alpha$-particle induced reactions with N, Ne and Si were deduced from the line integrals of the 13 background-subtracted spectra. In most cases the lines were isolated and a simple integration of line counts with subtraction of the Compton background was sufficient. For the 5.105-MeV line of $^{14}$N superposition with the second escape peak of the 6.129-MeV line had to be taken into account for proton irradiations with N$_2$O gas and for alpha-particle irradiations with N$_2$O and N$_2$. Lines that were not clearly standing out or sitting above a complicated background were not considered for analysis. We concentrated furthermore on lines produced by inelastic scattering.  

Uncertainties on the differential cross section data include the statistical uncertainty, and uncertainties due to background subtraction, dead-time correction and relative detection efficiency. The uncertainties of the dead-time correction and relative efficiencies were estimated to be smaller than 3\% and 5\%, respectively, the latter one depending on the $\gamma$-ray line energy. For the background subtraction we made, for each line integral, an estimation of the minimal and maximal background and took as the corresponding uncertainty half of the difference. All uncertainties were added quadratically.

In each clover detector, the crystals were arranged such that we had only two independent angles with respect to the beam direction. The data of the two crystals at the same angle were then averaged for the differential cross section. The $\gamma$-ray production cross sections were obtained by Legendre-polynomial fits to the differential cross sections:

\begin{equation}
\frac{d\sigma}{d\Omega}(\Theta) ~ = ~ \sum_{l=0}^{l_{max}}
a{_l}~ Q_{l}~ P_{l}(cos\Theta)~~~ (l~ even) 
\end{equation}

\noindent with $l_{max}$ being the smaller of twice the $\gamma$-ray transition multipolarity or twice the spin of the emitting state. Q$_l$ are attenuation coefficients, which were determined with the help of Geant simulations of the detection setup. The coefficient a$_0$ gives directly the angle-integrated cross section: $\sigma$ = 4$\pi$a$_0$.  

\subsubsection{Reactions with nitrogen}

We determined only cross sections for reactions with the major isotope $^{14}$N which represents more than 99.6\% of the isotopic abundance. Even at energies where spallation of $^{15}$N may produce $^{14}$N lines, all cross sections can safely be attributed to reactions with $^{14}$N within the given cross section uncertainties. 

In proton reactions we determined cross section excitation functions for 11 lines of $^{14}$N produced by inelastic scattering. They are from transitions of the first 8 excited states in the range E$_x$ = 2.313 - 7.029 MeV. Actually, we have at least one transition for each of these states, except for the J$^{\pi}$=0$^-$, 4.915 MeV state. The only visible de-excitation $\gamma$-ray line of this state sits at the Compton edge of the relatively strong 5.105-MeV line and was difficult to analyse despite the Compton suppression by the BGO shields. A list of the lines is presented in table \ref{table_Nlist}.

The strongest line up to E$_p$ = 20 MeV is the 2.313-MeV line from the first excited state transition. Due to the spin 0 of the emitting state the $\gamma$-ray angular distribution is isotropic. As it was also a prominent line in the spectra well above background and with good statistics it was very useful to check our relative detector efficiencies and dead-time corrections. Typical reduced $\chi^2$'s in the fits of the $\gamma$-ray angular distributions were in the range 0.5-1. Above E$_p$ = 6.36 MeV the nucleus $^{14}$O can be produced by the (p,n) reaction. It beta-decays with a halflife of t$_{1/2}$ = 70.6 s to the first excited state of $^{14}$N with a branching of nearly 100\%. This emission happens at rest and adds a very narrow ($\approx$4 keV FWHM) delayed component to the 2.313-MeV line which otherwise has a width of typically 50 keV for the prompt component from inelastic scattering  (see fig. \ref{figspec}). We did not try to separate the components and our cross section data thus represent the sum of prompt and delayed components. Decays happening after the irradiation runs were estimated and added to the line integrals. 
 
\begin{figure}
\includegraphics[height=.55\textheight]{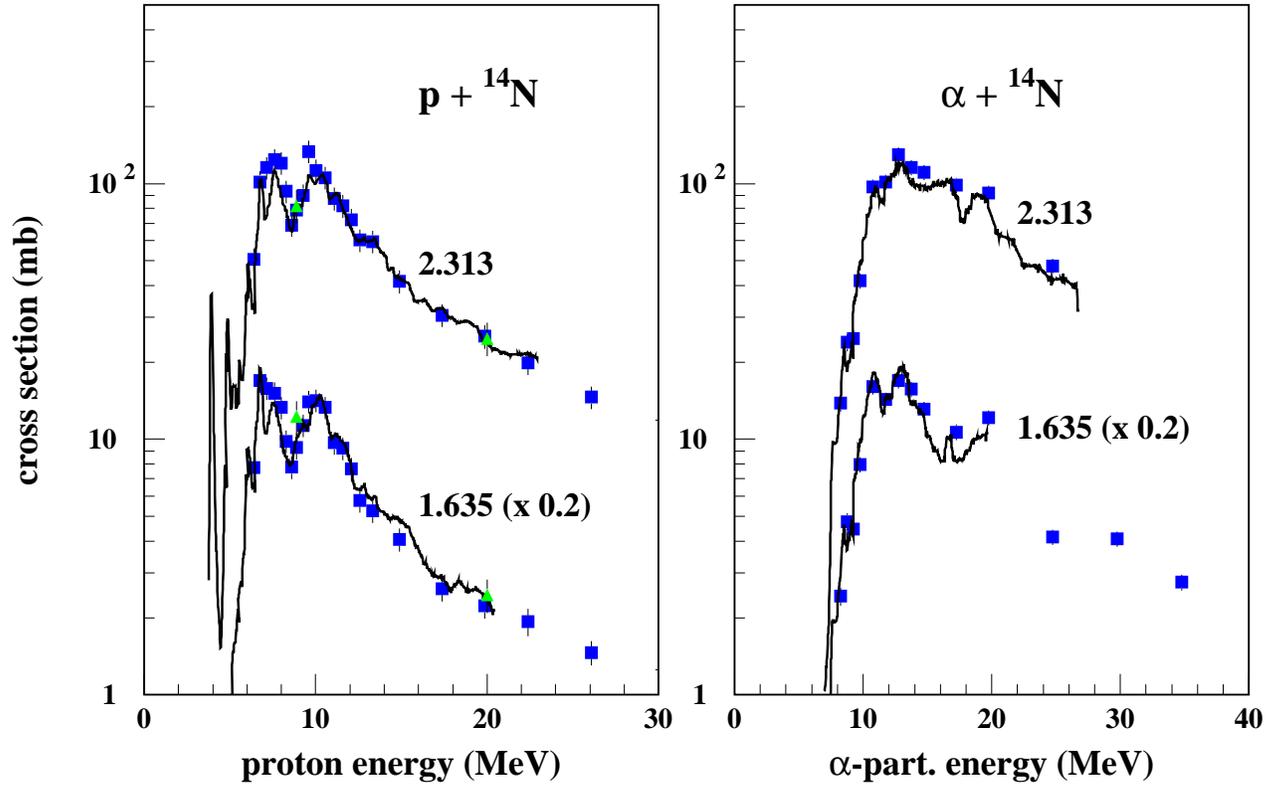}
\caption{(Color online) Cross section excitation functions of the $\gamma$-ray lines at 2.313 and 1.635 MeV in proton (left) and $\alpha$-particle reactions (right) with $^{14}$N. Blue squares are our data, green triangles are data of \cite{Lesko} and the full line data from the Washington lab. \cite{Dyerp} \cite{Dyera}. Cross sections for the 1.635-MeV line have been divided by 5 for better visibility.}
\label{figex1} 
\end{figure}

The transition from the second excited state produces the 1.635-MeV line with about 50\% of the 2.313-MeV line cross section and constitutes the second strongest line. For both the 2.313 and 1.635-MeV lines we determined cross section data for all proton irradiation runs in the range E$_p$ = 6.55 - 26.2 MeV. Cross sections in this energy range are available in \cite{Dyerp} and \cite{Lesko}. Comparison with their data shows a very good agreement of the excitation functions in the whole energy range (see fig. \ref{figex1}). 

Above E$_p$ = 20 MeV the strongest line is at $\sim$720 keV composed of the 0.728-MeV line  of $^{14}$N and the 0.718-MeV line of $^{10}$B. The latter one is from the $^{14}$N(p,p$\alpha$)$^{10}$B$^{\star}$ reaction whose laboratory threshold is E$_p$ = 13.2 MeV and which starts to dominate at 20 MeV. The 5.105-MeV line still has a cross section comparable to the 1.635-MeV line, the remaining lines are weaker, the maximum cross section does not exceed 25 mb. To our knowledge no other $\gamma$-ray data exist for these lines in the p + $^{14}$N reaction.

For the $\alpha$ + $^{14}$N reaction, we determined cross section excitation functions for  6 lines, from transitions of the 6 first excited states at E$_x$ = 2.313 - 5.834 MeV with the exception of the 4.915 MeV state. This time the two strongest lines are the 2.313-MeV line and the 5.105-MeV line with comparable cross sections followed by the 1.635-MeV line. Comparison with \cite{Seam} for the 2.313 and 1.635-MeV lines shows good agreement between both data sets. For the other 4 lines, no $\gamma$-ray data are available in the literature. A summary of all cross sections can be found in \cite{Benh}, and they are supplied in numerical form in \cite{epaps}.

\subsubsection{Reactions with neon}

For this element we considered reactions with two isotopes, $^{20}$Ne (isotopic abundance of 90.5\%) and $^{22}$Ne (9.25\%). The third stable isotope $^{21}$Ne has less than 0.3\% abundance and its contribution to $\gamma$-ray lines can be safely neglected. For consistency all cross sections are calculated with the total number of atoms even for cases where only one isotope can contribute to a given $\gamma$-ray line. A list of lines whose cross sections were measured in the neon irradiations is presented in table \ref{table_Nelist}.

$^{20}$Ne has only two excited states below its $\alpha$-particle emission threshold: 2$^+$, 1.634 MeV and 4$^+$, 4.248 MeV. We determined cross sections for the two lines from the de-excitation of these levels at 1.634 MeV and 2.614 MeV. The 1.634-MeV line in $\alpha$-particle induced reactions is blended with the 1.636-MeV line of $^{23}$Na  above E$_{\alpha}$ $\sim$6 MeV. We extracted furthermore data for the de-excitation line at 3.333 MeV from the third excited state  which cannot decay by $\alpha$-particle emission due to angular momentum selection rules. 

\begin{figure}
\includegraphics[height=.55\textheight]{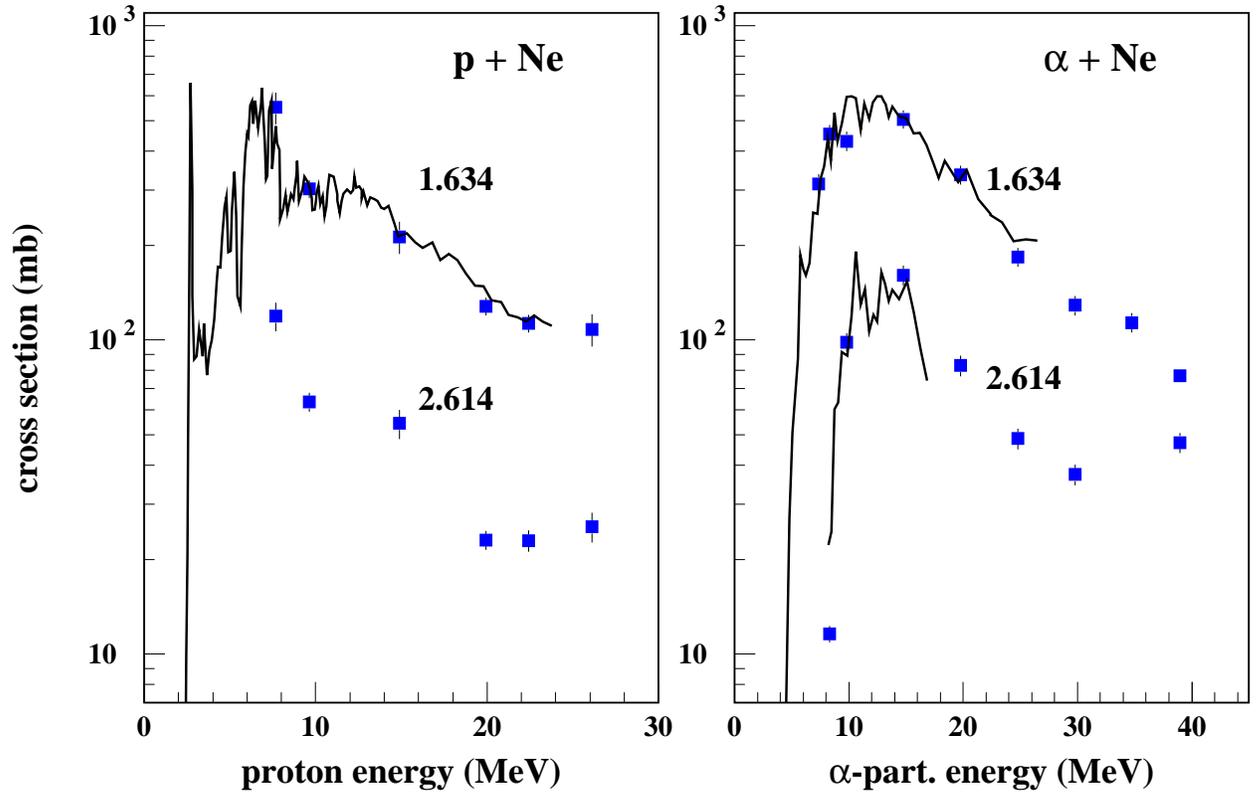}
\caption{(Color online) Cross section excitation functions of the $\gamma$-ray lines at 1.634 and 2.614 MeV in proton (left) and $\alpha$-particle reactions (right) with natural Ne. Squares are our data and the full line data from the Washington lab. \cite{Dyerp} \cite{Seam}.}
\label{figex2} 
\end{figure}

These lines are produced mainly by inelastic scattering off $^{20}$Ne. At projectile energies above $\sim$20 MeV reactions with $^{22}$Ne produce a small contribution not exceeding 10\% in our energy ranges. The 1.634-MeV line is also produced by the decay of $^{20}$Na (t$_{1/2}$ = 446 ms) above E$_p$ $\sim$15 MeV following the (p,n) reaction on $^{20}$Ne. Data for the 1.634-MeV line in proton reactions and for the 1.634 and 2.614-MeV lines in $\alpha$-particle reactions with Ne have already been measured by the Washington group \cite{Dyerp,Seam}. There is good agreement between their and our data sets (see fig. \ref{figex2}). No other $\gamma$-ray cross section data are known to us for the 3.333-MeV line in $\alpha$-particle and proton reactions and for the 2.614-MeV line in proton reactions. 

Concerning $^{22}$Ne, we have data for the lines at 1.275 MeV (2$^+$, 1.275 $\rightarrow$ 0$^+$, g.s.) and 2.083 MeV (4$^+$, 3.358 $\rightarrow$ 2$^+$, 1.275) for $\alpha$-particle reactions. These are produced by inelastic scattering off $^{22}$Ne below E$_{\alpha}$ = $\sim$15 MeV. Above that energy the $^{20}$Ne($\alpha$,2p)$^{22}$Ne reaction channel is open and provides certainly the major contribution above E$_{\alpha}$ = 25 MeV. For the proton reactions we extracted only data for the 1.275-MeV line, which is exclusively from proton-inelastic scattering off $^{22}$Ne. 

We also determined cross sections for the $^{23}$Na lines at 2.263 MeV (9/2$^+$, 2.704 $\rightarrow$ 5/2$^+$, 0.440) and 2.830 MeV (11/2$^+$, 5.534 $\rightarrow$ 9/2$^+$, 2.704) in $\alpha$-particle reactions above E$_{\alpha}$ = 15 MeV. They are completely dominated by the $^{20}$Ne($\alpha$,p)$^{23}$Na reaction channel below E$_{\alpha}$ = 30 MeV, while the $^{22}$Ne($\alpha$,X)$^{23}$Na reaction may contribute above that energy. Finally, we have cross sections for the 6.129-MeV line of $^{16}$O from proton irradiations of Ne in the range E$_p$ = 15 - 26.2 MeV. At these energies the line is practically only produced by the $^{20}$Ne(p,p$\alpha$)$^{16}$O reaction. A summary of all cross sections can be found in \cite{Benh}, and they are supplied in numerical form in \cite{epaps}.

\subsubsection{Reactions with silicon}

As in the case of Ne, cross sections were calculated for the total number of atoms even in cases where only one isotope may contribute to a given $\gamma$-ray line.  We determined cross section data in three proton irradiations of Si at E$_p$ = 10, 15 and 20 MeV, most of them for lines of the major isotope $^{28}$Si (92.2\%). A list of lines whose cross sections were measured in these irradiatons is presented in table \ref{table_Silist}.

Actually, for each of the first 11 excited states of $^{28}$Si we have at least one de-excitation line thus providing a full determination of the excited-state population of $^{28}$Si up to E$_x$ = 7.933 MeV. The most important component is inelastic scattering off $^{28}$Si with a small possible contribution of $^{28}$P decay (t$_{1/2}$ = 268 ms) at E$_p$ = 20 MeV where the $^{28}$Si(p,n)$^{28}$P reaction channel is open. A contribution of proton reactions with $^{29}$Si at E$_p$ = 15 and 20 MeV is also possible but should be small because of its relatively low isotopic abundance (4.7\%) with respect to $^{28}$Si. 

The by far strongest line is the 1.779-MeV line from the de-excitation of the first excited state approaching 400 mb at 15 MeV. The Washington group measured the cross section excitation function for this line in the range E$_p$ $\sim$ 3 - 23 MeV \cite{Dyerp}. The agreement between both data sets is very good at E$_p$ = 10 and 20 MeV but at E$_p$ = 15 MeV the cross sections disagree by more than 30\% which is outside the cross section uncertainties. We do not know what has caused this discrepancy. A usual measurement error like an incomplete charge collection seems unlikely in our experiment because the charge measurements in the two Faraday cups upstream and downstream of the target agreed to better than 10\% for this irradiation run. We decided anyhow to renormalize all our cross sections at 15 MeV because the cross section systematics for the 1.779-MeV line is without doubt in favor of the Washington data set. 

\begin{figure}
\includegraphics[height=.55\textheight]{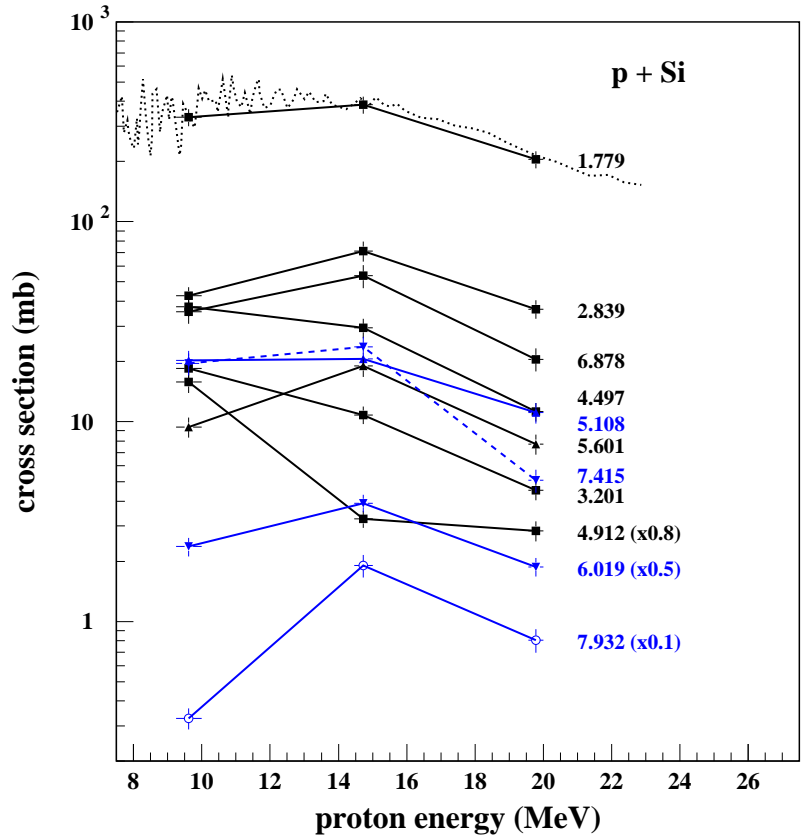}
\caption{(Color online) Cross section excitation functions of $\gamma$-ray lines from the strongest transitions of the first 11 excited states of $^{28}$Si. Symbols connected with lines are our data and the dotted line shows the Washington data for the 1.779-MeV line \cite{Dyerp}. }
\label{figex3} 
\end{figure}

The three next strongest lines with cross sections around 50 mb are the 2.839-MeV line from the de-excitation of the second excited state, the 4.497-MeV line from the 4$^{th}$ excited state  and the 6.878-MeV line of the 6$^{th}$ excited state. The other 9 lines have modest cross sections not exceeding 30 mb. A subset of our cross section data for $^{28}$Si lines can be found in fig. \ref{figex3} together with the Washington data \cite{Dyerp}.

For $^{29}$Si we could extract cross sections for 4 lines corresponding to de-excitation transitions  of the first 4 excited states. These lines are produced by inelastic scattering off $^{29}$Si and at E$_p$ = 15 MeV and 20 MeV also by reactions with $^{30}$Si. The cross sections are similar at these energies for both isotopes of similar isotopic abundance (4.7\% for $^{29}$Si and 3.1\% for $^{30}$Si). The decay of $^{29}$P proceeds with 98.3\% to the ground state of $^{29}$Si and is therefore not important for the studied $\gamma$-ray lines. 

We determined furthermore cross sections for two lines of $^{30}$Si, originating from the first two excited states. Here, only inelastic scattering off $^{30}$Si can contribute, the decay of $^{30}$P going to more than 99.9\% to the ground state of $^{30}$Si. 
 
Except for the 1.779-MeV line of $^{28}$Si, no other $\gamma$-ray cross section data for proton reactions with Si are published in the considered energy range. A summary of all cross sections can be found in \cite{Benh}, and they are supplied in numerical form in \cite{epaps}.

\section{Comparison with reaction code calculations}

Applications of measured $\gamma$-ray cross sections in modeling of astrophysical phenomena like solar flares or galactic cosmic rays is not straightforward and usually requires some preparation of the data. In both cases cross section excitation functions are needed from reaction threshold to energies of typically hundred MeV per nucleon for solar flares and to hundreds of MeV per nucleon for cosmic rays. A sufficiently complete set of experimental data to define the whole excitation function in the required energy range exists for practically no $\gamma$-ray line. Some theoretical input like nuclear reaction calculations is therefore essential for the modeling of $\gamma$-ray emissions in astrophysical sites. 

Modern nuclear reactions codes like EMPIRE \cite{Empire} and TALYS \cite{talys} have recently been used for such purposes. EMPIRE has been successfully employed reproducing cross sections of strong lines in proton and $\alpha$-particle reactions with $^{12}$C, $^{16}$O, $^{24}$Mg and Fe measured at the Orsay tandem \cite{tan2002}. Both codes have been used for the calculation of delayed $\gamma$-ray lines emitted by radioactive ions produced during solar flares \cite{TKKM}. TALYS has been chosen to complement cross section evaluations in the latest version of the $\gamma$-ray cross section database for astrophysical applications \cite{MKKS}. There, in particular, many cross section excitation functions of lines produced in $\alpha$-particle reactions have been updated to include an important contribution of fusion-evaporation reactions like ($\alpha$,pnd) and ($\alpha$,2p2n) which produce the same lines as inelastic scattering ($\alpha$,$\alpha$') and which had been neglected in the former versions. Another important point for calculations is the quasi-continuum of thousands of weak lines produced in astrophysical sites for which no experimental cross sections exist.

We chose to compare TALYS predictions with our data for $^{14}$N, Ne and Si lines. TALYS is a user-friendly code providing nuclear reaction calculations for $\gamma$, n, p, d, t, $^{3}$He and $\alpha$-particle induced reactions in the energy range E$_{lab}$ = 1 keV - 250 MeV \cite{talys}. The major reaction mechanisms in this range are included, otherwise TALYS uses comprehensive libraries for the optical potentials, deformations, discrete levels and level densities, masses etc. For most of the nuclear data the code offers a choice of values derived from experiment and from different theoretical models. The user can also input his own values by a set of keywords or simply edit the libraries. 

A first comparison was simply done with calculations using the default values of TALYS which have shown good agreement with experimental data for lines from transitions of the first few excited levels in abundant nuclei from C to Fe \cite{MKKS}. For most of the $\gamma$-ray lines here the agreement in the absolute cross section values is similar, typically within 50\%. For a few lines the difference between experimental and calulated values reaches a factor of two. The energy dependence is generally also reasonably reproduced with a notable exception for some lines of $^{14}$N in proton-induced reactions. 

\subsection{Reactions with $^{14}$N}

\begin{figure}
\includegraphics[height=.55\textheight]{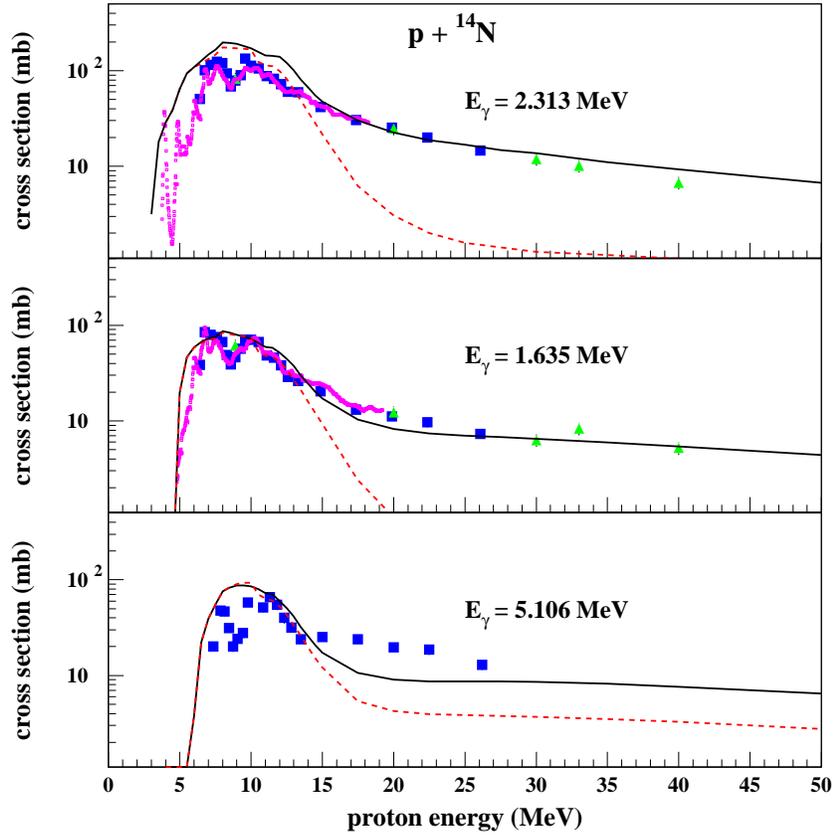}
\caption{(Color online) Cross section excitation function of the 2.313, 1.635 and 5.105-MeV lines for proton-inelastic scattering off $^{14}$N. The 2.313-MeV line production includes also the radioactive decay of $^{14}$O following the $^{14}$N(p,n)$^{14}$O reaction. Experimental data are from this experiment (blue squares), Lesko et al. \cite{Lesko} (green triangles) and Dyer et al. \cite{Dyerp} (small magenta squares). The dashed red line shows the TALYS calculation with default values, the full black line TALYS with modified deformation parameters.}
\label{fig_pn1635} 
\end{figure}

The most striking and clear deviation of a calculated energy dependence from measurement is found for the 2.313-MeV and 1.635-MeV lines in proton-induced reactions. In figure \ref{fig_pn1635} one can observe that TALYS with default values reproduces the bump of the excitation function up to E$_p$ = $\sim$15 MeV but falls severely below the experimental values above that energy. This is a clear indication that the compound nucleus component which usually dominates below 15 MeV is reasonably predicted while the direct reaction component expected to dominate above that energy is much too small in TALYS. In the latter mechanism the largest contribution to the 1.635-MeV line is from inelastic scattering to the second excited state of $^{14}$N at 3.948 MeV which has a nearly 100\% branching to the 2.313 MeV state. The cross section in the direct reaction mechanism is approximately proportional to the square of the deformation parameter $\beta_2$ for that state. In the present version of TALYS the default calculation is done with a $\beta_2$-value of practically zero. 

We introduced this state into the deformation data base in TALYS which defines the levels belonging to collective bands, their coupling and the deformation parameters. The 3.948-MeV state was included as the second member of the ground state rotational band, and the deformation parameter $\beta_2$ of the band was adjusted in order to reproduce the experimental data for the 1.635-MeV line above E$_p$ = 15 MeV. As starting parameter the value calculated from the B(E2) of the transition 3.948 MeV $\rightarrow$ g.s. (see e.g. \cite{Satchler}) was used. Additionally, we adjusted the deformation parameters $\beta_3$ and $\beta_2$ of the states at 5.834 and 7.029 MeV, respectively, in order to  improve the calculations for other lines, in particular the ones at 0.728 and 5.105 MeV. 

The values which are reported in table \ref{table_ndef} are close to the parameters extracted from analyses of inelastic proton scattering angular distributions (see e.g. \cite{Blanpied,Hansen}). The results for the 1.635-MeV line are shown in figure \ref{fig_pn1635}. One clearly observes a substantial improvement above E$_p$ $\sim$ 15 MeV with good agreement between calculated and measured values with adjusted $\beta_2$ parameter. A similar improvement can be seen in the same figure for the 2.313-MeV line which above 15 MeV is strongly produced by the cascade following inelastic scattering to the 3.948-MeV level. The improvement for the 1.635-MeV line therefore quite naturally improves also the agreement between calculation and measurement for this line. 

A second mechanism contributing significantly to the 2.313-MeV line is radioactive decay of $^{14}$O, following the $^{14}$N(p,n)$^{14}$O reaction. This contribution has been measured in our energy range in  several experiments, where globally the form of the cross section excitation functions agree in the different data sets while for the absolute normalization essentially two groups appear (see e.g. EXFOR \cite{EXFOR}). One group with maximum cross section at $\sim$10 MeV of about 100 mb, and one group with about 10 mb. From our spectra of the proton irradiation  at E$_p$ = 10.23 MeV a maximum contribution of 30\% to the 2.313-MeV line can be attributed to the $^{14}$N(p,n)$^{14}$O reaction (see fig. \ref{figspec}). This yields an upper limit of 35 mb to the (p,n) reaction, which is in complete disagreement with the excitation functions reaching 100 mb cross section at 10 MeV. We therefore choose to rely on the Washington data for the $^{14}$N(p,n)$^{14}$O reaction \cite{Dyerp}, which agree with \cite{Kitwanga,Nozaki,Kovacs}.

The TALYS calculation with default values overpredicts severely the $^{14}$N(p,n)$^{14}$O reaction by a factor of approximately 10 with respect to the Washington data. We were able to approach the measured values for the (p,n) reaction only with a change in the preequilibrium calculation (taking the multi-step direct/compound calculation instead of the default exciton model). The overall agreement of the TALYS calculation with measurement for the sum of inelastic scattering off $^{14}$N and decay of $^{14}$O is finally reasonable, although the cross section for the bump around E$_p$ = 10 MeV is still overestimated. 

The third line shown in fig. \ref{fig_pn1635} is the 5.105-MeV line. The cross section increase in the TALYS calculation with modified deformation parameters above E$_p$ = 15 MeV is due to the adjustment of the deformation parameter $\beta_3$ of the 5.834-MeV state which has a strong branching to the 5.106-MeV state. It is, however, not completely sufficient to reproduce the data above E$_p$ = 15 MeV. A still larger $\beta_3$ would overpredict the 0.728-MeV line cross section. The three lines shown here represent a typical sample for the degree of agreement between calculation and measurement, which for proton-induced reactions with $^{14}$N ranges from reasonable to moderate. A complete description of calculations can be found in \cite{Benh}. A better description of the cross section excitation functions probably requires a detailed study of the reaction mechanisms and the optical potentials for the incoming channel p + $^{14}$N and different outgoing channels like n + $^{14}$O and $\alpha$ + $^{11}$B. 

The default parameters in TALYS for these potentials are from global optical potentials which do not take into account the particularities of light nuclei. For protons and neutrons the default potential is the one of \cite{KonDel}, deduced in a comprehensive analysis of nucleon elastic scattering and reaction data with nuclei in the mass range 24 $\leq$ A $\leq$ 209. Potentials for the heavier particles d, t, $^{3}$He and $^4$He are obtained by a simplified folding approach of the neutron- and proton potentials (for more details see documentation of \cite{talys}). For light nuclei a detailed study of the optical potential in a large energy range has been done for p + $^{12}$C \cite{Meigooni}.  With this potential a good description of measured 4.438-MeV line shapes and the cross section excitation function in proton reactions with $^{12}$C has been obtained in optical-model calculations \cite{lshape,tan2002}. It is possible that a similar study for proton and $\alpha$-particle reactions with $^{14}$N could furthermore improve the calculations of $\gamma$-ray cross sections for this nucleus.

\begin{figure}
\includegraphics[height=.55\textheight]{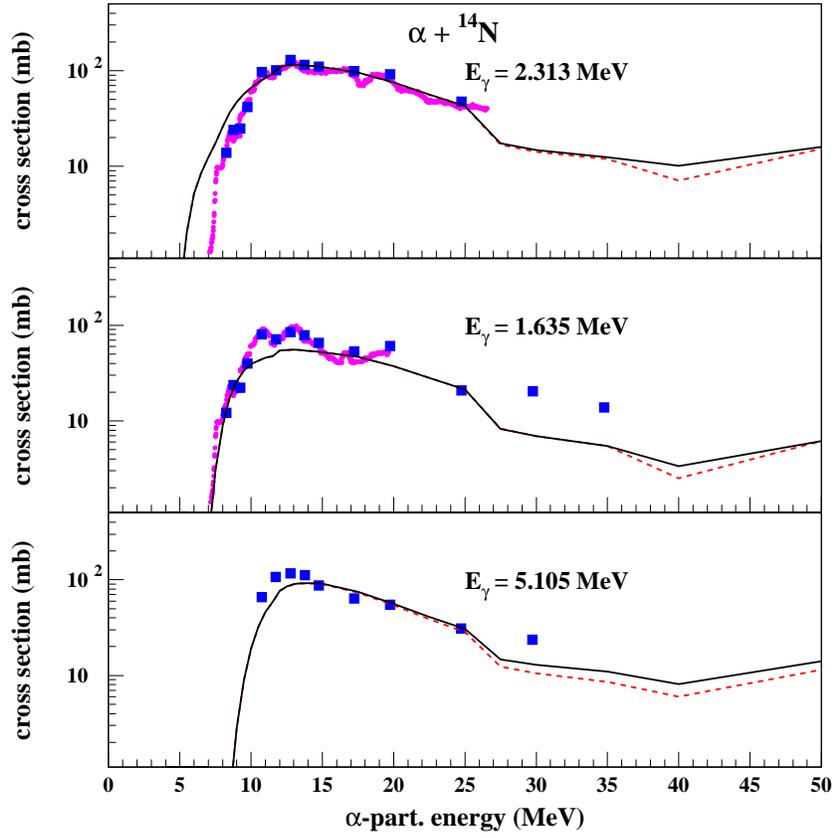}
\caption{(Color online) The same as figure \ref{fig_pn1635} except for $\alpha$-particle reactions with $^{14}$N.}
\label{fig_an1635} 
\end{figure}

In fig. \ref{fig_an1635} we show cross section excitation functions for the same three lines produced in $\alpha$-particle reactions with $^{14}$N. Here the adjustment of the deformation parameters has less impact on the cross sections because the direct reaction mechanism is nowhere dominating in the studied energy range. The reproduction of the experimental $\gamma$-ray line data is maybe slightly better than for proton reactions, which comes somehow as a surprise because $\alpha$-particle optical potentials are believed to be less reliable than the ones for protons. It may  point to the fact that there is a specific problem in the reaction mechanism for protons with $^{14}$N suggested also by the large overestimation of the $^{14}$N(p,n)$^{14}$O reaction cross section in the default TALYS calculation.

\subsection{Reactions with Ne and Si}

We concentrated our studies to $\gamma$-ray lines of the two neon isotopes, $^{20}$Ne, $^{22}$Ne and the main silicon isotope  $^{28}$Si. The first three levels of these isotopes have the typical spin-parity sequence: 0$^+$, 2$^+$, 4$^+$ of a collective ground-state vibrational or rotational band. In the default TALYS calculation a rotational coupling is considered for these levels with deformation parameters $\beta_2$ and $\beta_4$. Predictions here are already within typically 30\% of experimental data with a reasonable reproduction of the energy dependence for most of the $\gamma$-ray lines. We adjusted only slightly these deformation parameters to reproduce the cross section excitation functions of the $\gamma$-ray lines from the 4$^+$ - 2$^+$ and $2^+$ - $0^+$ transitions between these levels. 

For the two neon isotopes the same deformation parameters were used, considering $^{22}$Ne as a deformed $^{20}$Ne core plus two neutrons not contributing to the deformation. However, different parameters were necessary to reproduce the cross sections for proton and $\alpha$-particle reactions with Ne. The present parameters $\beta_2$ and $\beta_4$ for proton reactions yield very close deformation lengths ($\beta_L$R) compared to the values found in coupled-channel analyses of proton inelastic scattering angular distributions (see e.g. \cite{Swin, Swin3}). The values for $\alpha$-particle reactions, however, differ quite substantially from the values for proton reactions. Such a behaviour is certainly an indication that the optical model potentials for proton or most probably $\alpha$-particle reactions with $^{20}$Ne and $^{20}$Ne could need some improvement as it was the case for reactions with $^{14}$N.  The deformation parameters  are listed in table \ref{table_nedef}.

\begin{figure}
\includegraphics[height=.55\textheight]{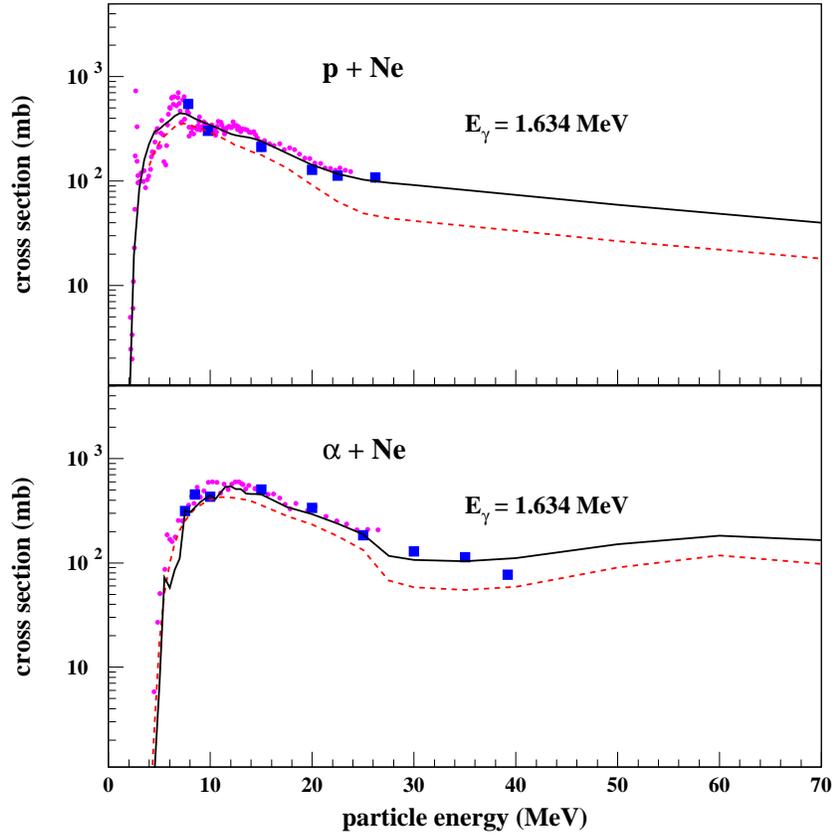}
\caption{(Color online) Cross section excitation functions of the 1.634-MeV line of $^{20}$Ne in proton and $\alpha$-particle reactions with Ne. Experimental data are from this experiment (blue squares), Dyer et al. \cite{Dyerp} for proton and Seamster et al. \cite{Seam} for $\alpha$-particle reactions (magenta circles). The dashed red line shows the TALYS calculation with default values, the full black line TALYS with modified deformation parameters.}
\label{fig_ne1634} 
\end{figure}

\begin{figure}
\includegraphics[height=.55\textheight]{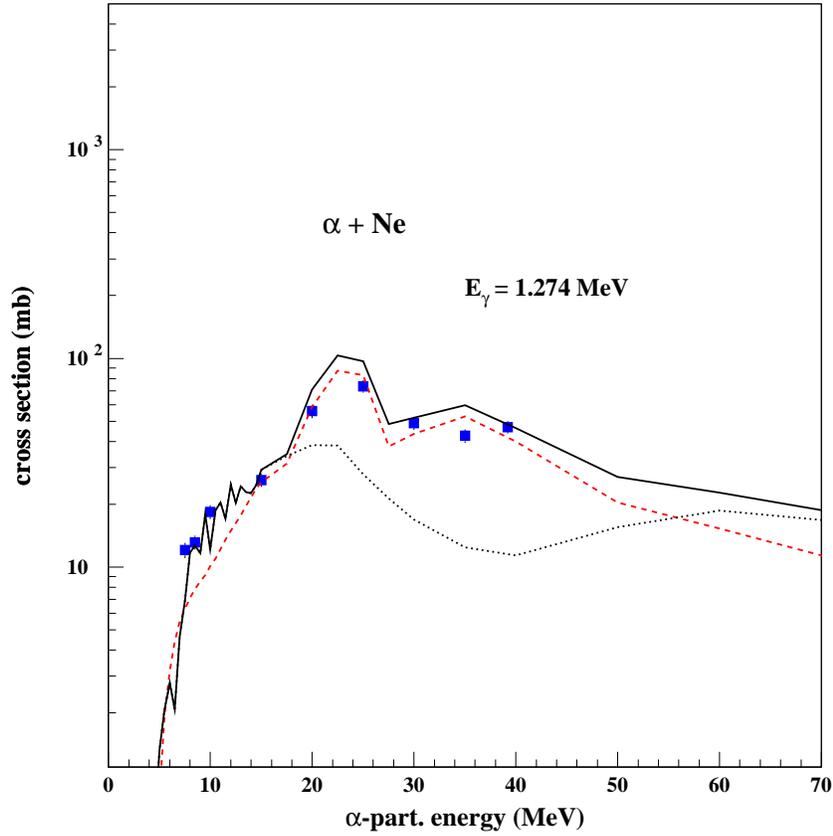}
\caption{(Color online) Cross section excitation functions of the 1.275-MeV line of $^{22}$Ne in $\alpha$-particle reactions with Ne. Shown are present experimental data (blue squares), the TALYS calculation with default values (dashed red line) and adjusted deformation parameters (full black line) and the inelastic scattering component $^{22}$Ne($\alpha$,$\alpha\gamma$)$^{22}$Ne (dotted black line). The dominating component in the range E$_{\alpha}$ = $\sim$20-50 MeV is from the $^{20}$Ne($\alpha$,2p$\gamma$)$^{22}$Ne reaction.}
\label{fig_ne1275} 
\end{figure}

\begin{figure}
\includegraphics[height=.55\textheight]{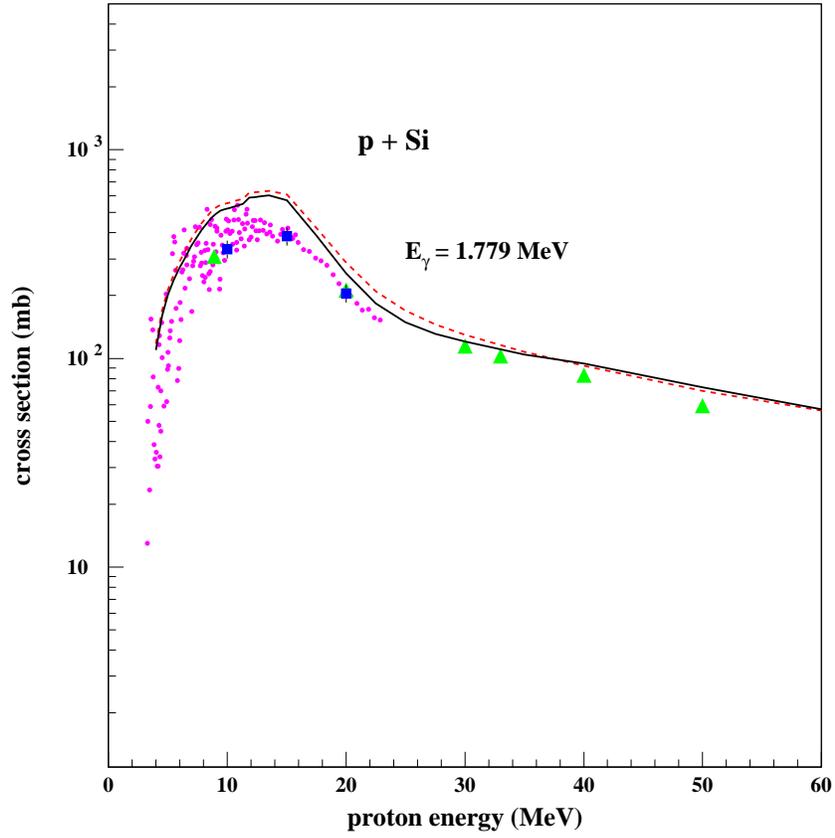}
\caption{Color online) Cross section excitation functions of the 1.779-MeV line of $^{28}$Si in proton reactions with Si. Experimental data are from this experiment (blue squares), Dyer et al. \cite{Dyerp} (magenta circles) and Lesko et al. \cite{Lesko} (green triangles). The dashed red line shows the TALYS calculation with default values, the full black line TALYS with adjusted deformation parameters.}
\label{fig_si1779} 
\end{figure}

For all $\gamma$-ray lines, contributions from reactions with all possible isotopes and of delayed radioactivity as explained above have been taken into account in the calculations. It resulted in a globally good agreement in absolute magnitude and energy dependence of the cross sections for these lines in proton- and $\alpha$-particle reactions with neon isotopes and in proton reactions with $^{28}$Si. Figures \ref{fig_ne1634} and \ref{fig_si1779} show the results for the strongest lines of $^{20}$Ne at 1.634 MeV and $^{28}$Si at 1.779 MeV which are dominated by inelastic scattering. The strongest line of $^{22}$Ne at 1.275 MeV has also a significant component from fusion-evaporation with $^{20}$Ne in $\alpha$-particle reactions as illustrated in fig. \ref{fig_ne1275}. Except for the 3.333-MeV line of $^{20}$Ne, the other lines produced in reactions with neon are reproduced with about the same quality as the lines presented in figures \ref{fig_ne1634} and \ref{fig_ne1275}. Calculation and experiment for the 3.333-MeV line differ by a factor of $\sim$2. 

With the slightly adjusted deformation parameters $\beta_2$ and $\beta_4$ of the ground-state band in $^{28}$Si, the calculations for the 1.779-MeV and 2.839-MeV lines agree to better than 20\% with the experimental data. Both parameters give deformation lengths compatible with the range found in coupled-channel or DWBA analyses of proton inelastic scattering angular distributions (see, e.g., \cite{Swin2,Lisantti,Knoe, Swin3} and references therein). The TALYS calculations for the other 13 lines of $^{28}$Si show reasonable to good agreement with measured data at E$_p$ = 10 and 15 MeV. We obtained furthermore good agreement at E$_p$ = 20 MeV for nearly all lines by adjusting slightly the deformation parameter $\beta_3$ for the J$^{\pi}$=3$^-$, 6.879 MeV state and by adding three high-lying 2$^+$ states to the collective level coupling scheme. The deformation parameters are listed in table \ref{table_sidef}. 

\begin{figure}
\includegraphics[height=.55\textheight]{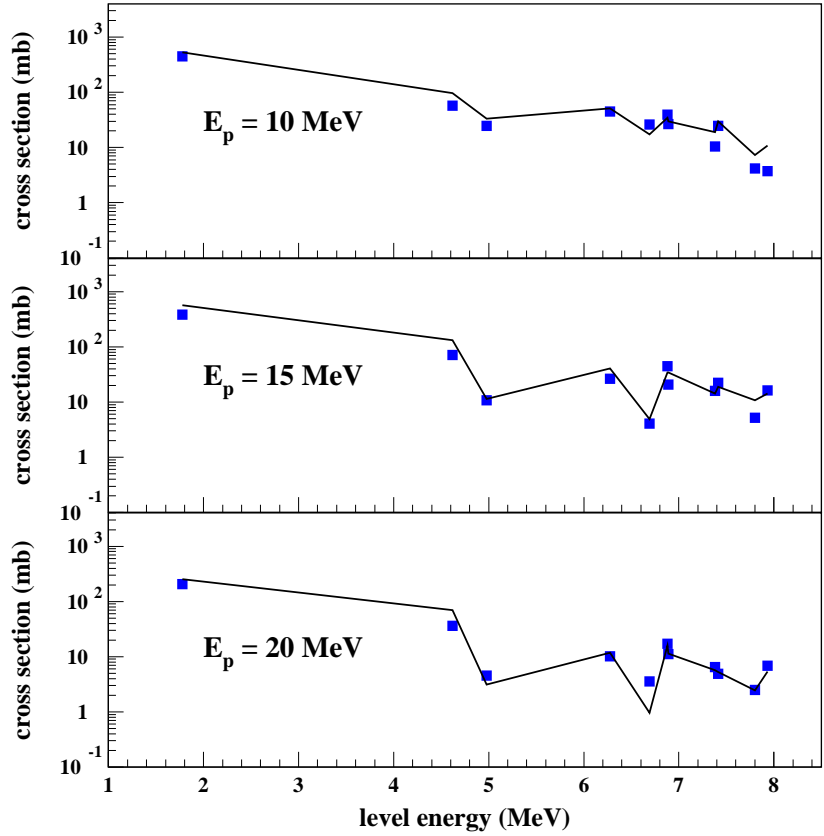}
\caption{$\gamma$-ray emission cross sections of the first 11 excited states of $^{28}$Si in p + Si reactions as a function of excitation energy E$_x$ at three different proton energies. Symbols represent our data and the full line the results of TALYS calculations with adjusted deformation parameters.}
\label{fig_siall} 
\end{figure}

Figure \ref{fig_siall} shows a comparison of measured and calculated  cross sections for the total $\gamma$-ray emission of the first 11 excited states of $^{28}$Si. They have been obtained with the help of the known $\gamma$-ray branchings (taken from NNDC \cite{nudat}); in cases where there are data for several transitions of a given nuclear state, a weighted mean was adopted. The observed good agreement between calculations and experiment at all proton energies illustrates the globally correct description of the population of excited states of $^{28}$Si up to E$_x$ = 8 MeV for the compound-nucleus and the direct reaction mechanisms in TALYS.  

\subsection{Conclusions}

With the present experiment all important nuclei from C to Fe for the $\gamma$-ray line emission induced by energetic protons and $\alpha$ particles in astrophysical sites have now been studied at the Orsay tandem accelerator. Due to the use of a gas target we have collected an abundant set of high-quality line shapes for the  the 6.129-MeV line of $^{16}$O. For the $^{16}$O line data exist in a neat energy coverage from threshold to 26.2 MeV for proton reactions and from threshold to 39.2 MeV for $\alpha$-particle reactions with $^{16}$O.  Together with the already existing data on the 4.444-MeV line of $^{12}$C we shall eventually have accurate line-shape calculations for two of the most prominent lines in solar flares.  

13 new $\gamma$-ray cross section excitation functions have been measured for proton and $\alpha$-particle reactions with $^{14}$N, 8 with Ne and 19 for proton reactions with Si. There is now in particular a nearly complete coverage for excited states of $^{14}$N and $^{28}$Si up to  E$_x$ = $\sim$8 MeV by $\gamma$-ray data. It has been shown that these data are very useful for testing nuclear reaction code calculations. The data for the 14 lines of $^{28}$Si allowed us to complete the coupling scheme of collective levels and adjust the deformation parameters. We finally obtained a good reproduction of the population of the first 11 excited states of $^{28}$Si for a range of proton energies extending from dominating compound-nucleus to dominating direct reaction mechanisms. This gives some confidence for the extrapolation to higher proton energies where no $\gamma$-ray data are available. A similar success could be obtained for the few $\gamma$-ray emitting levels of $^{20}$Ne. For the other two very abundant nuclei in the mass range A $\geq$ 20 - $^{24}$Mg and $^{56}$Fe - a great amount of $\gamma$-ray data is available in the literature in particular from the Washington and Orsay tandem accelerators to proceed in the same way.   

With the $\gamma$-ray data for $^{14}$N we could propose a new coupling scheme and adjusted deformation parameters for the included levels. It resulted in a dramatic improvement in the description of the energy dependence for several $\gamma$-ray lines in proton reactions with that nucleus, even if the final overall agreement of calculations with measured data could still be improved. This illustrates the need of a specific treatment of light nuclei in nuclear reaction calculations by studying for example in detail the optical potentials for proton and $\alpha$-particle reactions. At least three such nuclei - $^{12}$C, $^{14}$N and $^{16}$O - are important for astrophysical sites like solar flares and galactic cosmic-ray interactions. We have now a comprehensive set of $\gamma$-ray data for all three of them which should help to establish reliable calculations for the $\gamma$-ray line emission in proton and $\alpha$-particle induced nuclear reactions.

\newpage

\newpage

\newpage

\newpage

\newpage

\newpage

\newpage

\newpage

\newpage

\newpage

\newpage

\newpage

\newpage

\begin{table}
\caption{Gas densities and typical beam energy losses in the different irradiation runs.}

\begin{tabular}{cccc}
Reaction  &   $\rho$ (mg/cm$^2$)  & E$_{lab}$ (MeV)   & $\Delta$E$_{lab}$/2 (MeV)  \\
 \hline \\
 p + N$_2$O & 5.20   & 6.55         & 0.15   \\
         & 7.50      & 6.95 - 7.35  & 0.20 \\
         & 9.92      & 7.84         & 0.24 \\
         & 6.50      & 8.14 - 9.04  & 0.15 \\
         & 8.87      & 9.44 - 9.78  & 0.18  \\
         & 11.11      & 10.23       & 0.22 \\
         & 15.63      & 10.83       & 0.29 \\
         & 13.97      & 11.33 - 12.33       & 0.24 \\
         & 15.16      & 12.83       & 0.25 \\
         & 11.89      & 13.5       & 0.19 \\
 $\alpha$ + N$_2$O     & 0.92     & 7.0 - 9.0  & 0.24   \\
         & 1.08       & 9.5 - 11.0  & 0.25   \\
         & 1.29       & 12.0 - 14.0  & 0.25   \\
p +  N$_2$  & 10.81   & 15.0 - 25.0  & 0.10 - 0.16  \\
$\alpha$ +  N$_2$     & 1.43    & 15.0   & 0.25  \\
        & 1.77      & 17.5 - 22.5  & 0.26   \\
        & 2.29      & 25.0 - 30.0  & 0.25   \\
        & 2.92      & 35.0 - 39.2  & 0.25   \\
p +  Ne    & 7.72   & 7.5 - 26.2  & 0.07 - 0.18  \\
$\alpha$ +  Ne        & 1.41  & 10.0 - 20.0  & 0.18 - 0.30  \\
        & 2.40       & 25.0 - 30.0  & 0.23   \\
        & 3.20       & 35.0 - 39.2  & 0.25   \\
 p + CO$_2$ & 16.99   & 15.0 - 26.2  & 0.15 - 0.24  \\
$\alpha$ +  CO$_2$     & 1.68    & 15.0 - 20.0  & 0.27  \\
        & 2.26       & 22.5 - 30.0  & 0.27   \\
        & 2.97       & 35.0 - 39.2  & 0.25   \\

  \end{tabular} 
\label{table_runs}
\end{table}


\begin{table}
\caption{List of measured $\gamma$-ray line cross sections in reactions with N.}

\begin{tabular}{clll}
line energy (MeV) & transition & beam energy (MeV) & other data  \\
 \hline \\
0.728  & $^{14}$N 3$^-$ 5.834 $\rightarrow$ 2$^-$ 5.106$^a$ & (p) 7.35 - 26.2  &  \\
 & & ($\alpha$) 11.0 - 35.0 & \\  
1.635  & $^{14}$N 1$^+$ 3.947 $\rightarrow$ 0$^+$ 2.313 & (p) 6.55 - 26.2  & \cite{Dyerp} (3.8-23.0 MeV), \cite{Lesko} (8.9-40.0 MeV) \\
 & & ($\alpha$) 8.5 - 35.0 & \cite{Dyera} (7-20 MeV) \\
2.313  & $^{14}$N 0$^+$ 2.313 $\rightarrow$ 1$^+$ g.s. & (p) 6.55 - 26.2  & \cite{Dyerp} (3.8-23.0 MeV), \cite{Lesko} (8.9-40.0 MeV) \\
 & & ($\alpha$) 8.5 - 25.0 & \cite{Dyera} (7-27 MeV) \\
2.793  & $^{14}$N 2$^-$ 5.106 $\rightarrow$ 0$^+$ 2.313 & (p) 7.35 - 15.0  &  \\
 & & ($\alpha$) 11.0 - 30.0 & \\  
3.378  & $^{14}$N 1$^-$ 5.691 $\rightarrow$ 0$^+$ 2.313 & (p) 7.35 - 15.0  &  \\
 & & ($\alpha$) 15.0 - 35.0 & \\  
3.890  & $^{14}$N 1$^+$ 6.204 $\rightarrow$ 0$^+$ 2.313 & (p) 7.84 - 15.0  &  \\
5.105  & $^{14}$N 2$^-$ 5.106 $\rightarrow$ 1$^+$ g.s. & (p) 7.35 - 26.2  &  \\
 & & ($\alpha$) 11.0 - 30.0 & \\  
5.690  & $^{14}$N 1$^-$ 5.691 $\rightarrow$ 1$^+$ g.s. & (p) 7.35 - 8.14  &  \\
5.833  & $^{14}$N 3$^-$ 5.834 $\rightarrow$ 1$^+$ g.s. & (p) 7.35 - 8.14  &  \\
6.445  & $^{14}$N 3$^+$ 6.446 $\rightarrow$ 1$^+$ g.s. & (p) 7.84 - 8.44  &  \\
7.027  & $^{14}$N 2$^+$ 7.029 $\rightarrow$ 1$^+$ g.s. & (p) 8.44 - 26.2  &  \\
\end{tabular} 
\noindent
$^a$ composite line, other transitions: $^{10}$B 1$^+$ 0.718 $\rightarrow$ 3$^+$ g.s.;
\label{table_Nlist}
\end{table}


\begin{table}
\caption{List of measured $\gamma$-ray line cross sections in reactions with Ne.}

\begin{tabular}{clll}
line energy (MeV) & transition & beam energy (MeV) & other data  \\
 \hline \\
1.634  & $^{20}$Ne 2$^+$ 1.634 $\rightarrow$ 0$^+$ g.s. & (p) 7.84 - 26.2  &  \cite{Dyerp}  (2-23 MeV) \\
 & $^{20}$Ne$^a$ & ($\alpha$) 7.5 - 39.2 & \cite{Seam} (4.5-26.5 MeV) \\  
2.614  & $^{20}$Ne 4$^+$ 4.248 $\rightarrow$ 2$^+$ 1.634 & (p) 7.84 - 26.2  & \\
 & & ($\alpha$) 8.5 - 39.2 & \cite{Seam} (8-17 MeV) \\
3.333  & $^{20}$Ne 2$^-$ 4.967 $\rightarrow$ 2$^+$ 1.634 & (p) 7.84 - 26.2  & \\
 & & ($\alpha$) 15.0 - 20.0 &  \\
1.275  & $^{22}$Ne 2$^+$ 1.275 $\rightarrow$ 0$^+$ g.s. & (p) 7.84 - 26.2  &  \\
 & & ($\alpha$) 7.5 - 39.2 & \\  
2.083  & $^{22}$Ne 4$^+$ 3.358 $\rightarrow$ 2$^+$ 1.275 & ($\alpha$) 15.0 - 25.0  &  \\
2.263  & $^{23}$Na 9/2$^+$ 2.704 $\rightarrow$ 5/2$^+$ 0.440 & ($\alpha$) 15.0 - 35.0  &  \cite{Seam} (8-26 MeV)\\
2.830  & $^{23}$Na 11/2$^+$ 5.534 $\rightarrow$ 9/2$^+$ 2.704 & ($\alpha$) 15.0 - 25.0  &  \\
6.129  & $^{16}$O 3$^-$ 6.130 $\rightarrow$ 0$^+$ g.s. & (p) 15.0 - 26.2  & \cite{Dyerp} (17-23 MeV) \\
 & & ($\alpha$) 11.0 - 30.0 &  \\  
\end{tabular} 

$^a$ composite line, other transitions: $^{23}$Na 7/2$^+$ 2.076 $\rightarrow$ 5/2$^+$ 0.440;
\label{table_Nelist}
\end{table}


\begin{table}
\caption{List of measured $\gamma$-ray line cross sections in proton reactions with Si.}

\begin{tabular}{clll}
line energy (MeV) & transition & beam energy (MeV) & other data  \\
 \hline \\
1.658  & $^{28}$Si 3$^+$ 6.276 $\rightarrow$ 4$^+$ 4.618 & (p) 10.0 - 20.0  &  \\
1.779  & $^{28}$Si 2$^+$ 1.779 $\rightarrow$ 0$^+$ g.s. & (p) 10.0 - 20.0  &  \cite{Dyerp}  (3-23 MeV) \\
2.839  & $^{28}$Si 4$^+$ 4.618 $\rightarrow$ 2$^+$ 1.779 & (p) 10.0 - 20.0  &  \\
3.201  & $^{28}$Si 0$^+$ 4.980 $\rightarrow$ 2$^+$ 1.779 & (p) 10.0 - 20.0  &  \\
4.497  & $^{28}$Si 3$^+$ 6.276 $\rightarrow$ 2$^+$ 1.779 & (p) 10.0 - 20.0  &  \\
4.912  & $^{28}$Si 0$^+$ 6.691 $\rightarrow$ 2$^+$ 1.779 & (p) 10.0 - 20.0  &  \\
5.099  & $^{28}$Si 3$^-$ 6.879 $\rightarrow$ 2$^+$ 1.779 & (p) 10.0 - 20.0  &  \\
5.108  & $^{28}$Si 4$^+$ 6.888 $\rightarrow$ 2$^+$ 1.779 & (p) 10.0 - 20.0  &  \\
5.601  & $^{28}$Si 2$^+$ 7.381 $\rightarrow$ 2$^+$ 1.779 & (p) 10.0 - 20.0  &  \\
6.019  & $^{28}$Si 3$^+$ 7.799 $\rightarrow$ 2$^+$ 1.779 & (p) 10.0 - 20.0  &  \\
6.878  & $^{28}$Si 3$^+$ 6.879 $\rightarrow$ 0$^+$ g.s. & (p) 10.0 - 20.0  &  \\
7.380  & $^{28}$Si 2$^+$ 7.381 $\rightarrow$ 0$^+$ g.s. & (p) 10.0 - 20.0  &  \\
7.415  & $^{28}$Si 2$^+$ 7.416 $\rightarrow$ 0$^+$ g.s. & (p) 10.0 - 20.0  &  \\
7.932  & $^{28}$Si 2$^+$ 7.933 $\rightarrow$ 0$^+$ g.s. & (p) 10.0 - 20.0  &  \\
1.273  & $^{29}$Si 3/2$^+$ 1.273 $\rightarrow$ 1/2$^+$ g.s. & (p) 10.0 - 20.0  &  \\
1.794  & $^{29}$Si 5/2$^+$ 3.067 $\rightarrow$ 3/2$^+$ 1.274 & (p) 10.0 - 20.0  &  \\
2.028  & $^{29}$Si 5/2$^+$ 2.028 $\rightarrow$ 1/2$^+$ g.s. & (p) 10.0 - 20.0  &  \\
2.426  & $^{29}$Si 3/2$^+$ 2.426 $\rightarrow$ 1/2$^+$ g.s. & (p) 10.0 - 20.0  &  \\
1.263  & $^{30}$Si 2$^+$ 3.498 $\rightarrow$ 2$^+$ 2.235 & (p) 10.0 - 20.0  &  \\
2.235  & $^{30}$Si 2$^+$ 2.235 $\rightarrow$ 0$^+$ g.s. & (p) 10.0 - 20.0  &  \\
\end{tabular} 
\label{table_Silist}
\end{table}

\newpage

\begin{table}
\caption{Adjusted values of the deformation parameters for proton and $\alpha$-particle reactions with $^{14}$N in the TALYS calculations. 'R' means rotational coupling and 'D' stands for weakly coupled levels which are treated in DWBA. The reduced radii for the volume real and imaginary potentials were r$_{V,W}$ = 1.136 fm for both reaction channels.}

\begin{tabular}{ccccccc}
E$_x$ (MeV) & \hspace{0.2 cm} J$^{\pi}$ \hspace{0.2 cm} & coupling & \hspace{0.2 cm} $\beta_2$ \hspace{0.2 cm} & \hspace{0.2 cm} $\beta_{3}$ \hspace{0.2 cm} \\ 
 \hline \\
 0.000 & 1$^+$ & R & 0.25 & \\
 3.948 & 1$^+$ & R &   \\
 5.834 & 3$^-$ & D &  &  0.48 \\
 7.029 & 2$^+$ & D & 0.34 &   \\
\end{tabular} 
\label{table_ndef}
\end{table}

\newpage

\begin{table}
\caption{Adjusted values of the deformation parameters for proton and $\alpha$-particle reactions  with $^{20}$Ne and $^{22}$Ne in the TALYS calculations. The reduced radii for the volume real and imaginary potentials were r$_{V,W}$ = 1.155 and 1.159 fm for reactions with $^{20}$Ne and $^{22}$Ne, respectively. }

\begin{tabular}{ccccccc}
reaction & E$_x$($^{20}$Ne) (MeV) & \hspace{0.2 cm} J$^{\pi}$ \hspace{0.2 cm} & coupling & \hspace{0.2 cm} $\beta_2$ \hspace{0.2 cm} & \hspace{0.2 cm} $\beta_4$ \hspace{0.2 cm} & E$_x$($^{22}$Ne) (MeV) \\ 
 \hline \\
p + Ne & 0.000 & 0$^+$ & R & 0.47 &  0.28 & 0.000 \\
 &       1.635 & 2$^+$ & R &      &       & 1.275  \\
 &       4.248 & 4$^+$ & R &      &       & 3.348 \\
$\alpha$ + Ne & 0.000 & 0$^+$ & R & 0.67 &  0.35 & 0.000 \\
 &              1.635 & 2$^+$ & R &      &       & 1.275  \\
 &              4.248 & 4$^+$ & R &      &       & 3.348 \\
 \end{tabular} 
\label{table_nedef}
\end{table}

\clearpage

\newpage

\begin{table}
\caption{Adjusted values of the deformation parameters for proton reactions with $^{28}$Si in the TALYS calculations. 'R' means rotational and 'V' vibrational coupling.  The reduced radii for the volume real and imaginary potentials were r$_{V,W}$ = 1.170 fm. }

\begin{tabular}{cccccc}
 E$_x$ (MeV) & \hspace{0.2 cm} J$^{\pi}$ \hspace{0.2 cm} & coupling & \hspace{0.2 cm} $\beta_2$ \hspace{0.2 cm} & \hspace{0.2 cm} $\beta_4$ \hspace{0.2 cm} & \hspace{0.2 cm} $\beta_3$ \hspace{0.2 cm} \\ 
  \hline \\
 0.000 & 0$^+$ & R & -0.343 & 0.176 &   \\
 1.779 & 2$^+$ & R &        &       &   \\
 4.618 & 4$^+$ & R &        &       &   \\
 6.879 & 3$^-$ & V &        &       &  0.305 \\
 7.381 & 2$^+$ & V & 0.2    &       &  \\
 7.416 & 2$^+$ & V & 0.1    &       &  \\
 7.933 & 2$^+$ & V & 0.15   &       &  \\

\end{tabular} 
\label{table_sidef}
\end{table}

\end{document}